\documentclass[twocolumn,showpacs,preprintnumbers,amsmath,amssymb]{revtex4}
\usepackage{graphicx}
\usepackage{dcolumn}
\usepackage{bm}
\begin{document}

\preprint{UA-NPPS/04/2003}
\title{Observational Critical QCD}

\author{N.G. Antoniou} 
\email{nantonio@cc.uoa.gr}
\author{Y.F. Contoyiannis}
\author{F.K. Diakonos}
\email{fdiakono@cc.uoa.gr}
\author{G. Mavromanolakis}
\email{gmavroma@mail.cern.ch}
\affiliation{Department of Physics, University of Athens, GR-15771 Athens, 
Greece}

\date{\today}

\begin{abstract}
A detailed study of correlated scalars, produced
in collisions of nuclei and associated with the $\sigma$-field
fluctuations, $(\delta \sigma)^2= < \sigma^2 >$, at the QCD critical
point (critical fluctuations), is performed on the basis of a
critical event generator (Critical Monte-Carlo) developed in 
our previous work. The aim of this analysis is to reveal 
suitable observables
of critical QCD in the multiparticle environment of simulated events
and select appropriate signatures of the critical point, associated
with new and strong effects in nuclear collisions.
\end{abstract}


\maketitle
The existence of a critical point in the phase diagram of QCD, for
nonzero baryonic density, is of fundamental significance for our
understanding of strong interactions and so its experimental verification
has become an issue of high priority \cite{WRB}. For this purpose an 
extensive programme of
event-by-event searches for critical fluctuations in the pion sector
has already been started in experiments with heavy ions from SPS to
RHIC energies \cite{NA49}. In \cite{ACDKK} we have emphasized, however, 
that in order to reveal critical density fluctuations in multiparticle
environment, one has to reconstruct the $\sigma$-sector from pion
pairs ($\sigma \to \pi^{+} \pi^{-}$) and study correlations of
sigmas as a function of the invariant mass near the two -pion
threshold. In fact, the QCD critical point, if it exists, communicates
with a zero mass scalar field ($\sigma$-field) which at lower
temperatures ($T < T_c$) may reach the two-pion threshold and decay
in very short time scales owing to the
fact that its coupling to the two-pion system is very strong. Obviously,
the fundamental, underlying pattern of $\sigma$-field fluctuations,
built-up near the critical point by the universal critical exponents
of QCD \cite{ACDKK}, is phenomenologically within reach if and only if
the study of correlated sigmas, reconstructed near the two-pion
threshold, becomes feasible. In this Letter we perform a detailed
feasibility study of the observables related to the detection
of the QCD critical point in nuclear collisions. In order to proceed 
we summarize, first, the principles on which the behaviour
of a critical system of sigmas is based \cite{ACDKK,WRB}. 

(a) The geometrical structure  of the critical system in transverse
space (after integrating in rapidity) consists of $\sigma$-clusters 
with a fractal dimension $d_F=\frac{2(\delta-1)}{\delta+1}$
leading to a power law, $< \sigma^2> \sim \vert \vec{x} \vert^{-
\frac{4}{\delta+1}}$, for the $\sigma$-field fluctuations, within each
cluster ($\delta$ : isotherm critical exponent)

(b) In transverse momentum space the $\sigma$-fluctuations obey a
power law $< \sigma^2> \sim \vert \vec{p}_{\perp} \vert^{-\frac
{2(\delta-1)}{\delta+1}}$ leading to observable intermittent
behaviour of factorial moments: $F_2(M) \sim (M^2)^{\frac{\delta-1}
{\delta+1}}$ where $M^2$ is the number of $2D$ cells \cite{BialPes}.

(c) The density fluctuations of pion pairs $(\pi^{+}\pi^{-})$ with
invariant mass close to two-pion threshold $(2 m_{\pi})$ simulate
to a good approximation the sigma-field fluctuations, $(\delta \sigma)^2
\approx <\sigma^2>$, at the critical point, under the assumption that
the sigma mass reaches the two-pion threshold $(m_{\sigma} 
\stackrel{>}{\sim} 2 m_{\pi})$ at a freeze-out temperature 
close to the critical value.

(d) The QCD critical point belongs to the universality class of the
$3D$ Ising system in which $\delta \approx 5$.

On the basis of these principles and the fact that critical
clusters in the above universality class interact weakly \cite{ACDPRE}, 
one may construct a generator
of critical events (Critical Monte-Carlo: CMC) containing only
sigmas, correlated according to the above prescription \cite{ACDKK}. 
The numerical
experiment is completed by a self-consistent treatment of 
$\sigma$-reconstruction in the simulated events (after the process of decay of
sigmas to pions) and a comparative study between
the associated critical correlations-fluctuations and the corresponding
behaviour of a conventional Monte-Carlo (HIJING).

The input parameters of the simulation are 
the size of the system in rapidity ($\Delta$) and transverse
space ($R_{\perp}$) as well as the proper time scale ($\tau_c$) 
characteristic for the formation of the critical system. 
The CMC generator can be used to obtain
a large set of critical events. The analysis of these events
could serve as a guide for the search of the
critical point in the data of real $A+A$ processes in SPS 
and RHIC experiments. This will be shown in a more transparent
way in what follows.

We will investigate two sets of critical events. One set corresponding
to a rather small system with mean charged pion multiplicity per event
$<n_{ch}> \approx 40$ and another set describing a larger system 
($<n_{ch}>\approx 120$). Each data set consists of $10^5$ CMC generated
events. First we will consider the
properties of the small system. For the values of the input parameters
in this case we take $\Delta=6$ adapted to the SPS energies and 
$R_{\perp}=15~fm$. 
In order to reveal underlying critical fluctuations, at the level of
observation, one has first to perform factorial moment analysis in
small cells of the phase space \cite{BialPes}. We have chosen transverse
momenta for this analysis, the reason being that density fluctuations
in this plane are practically independent of the geometry of the
critical system \cite{ACDKK}. 
Applying factorial moment analysis to the 
transverse momenta $(p_x,p_y)$ of the negative pions in the sample
of the $10^5$ critical events we obtain for the second moment the
behaviour shown in Fig.~1. We observe a weak intermittency effect
($F_2 \sim M^{2 s_2}$ with $s_2 \approx 0.07$) much smaller than the 
expected to occur in the critical system ($s_2 = 2/3$). 
For comparison we show in the same plot the corresponding factorial
moment for the sigmas, before their decay, which follows closely the
theoretical prediction: $s_2 \approx 0.63$. The pions analysed in this
figure originate from the decay of the CMC sigmas using pure kinematics
and taking into account the isospin properties of the decay amplitude.
The mass of the decaying sigmas is treated as an almost uniformly 
distributed random variable \cite{ACDKK}.

\begin{figure}
\includegraphics{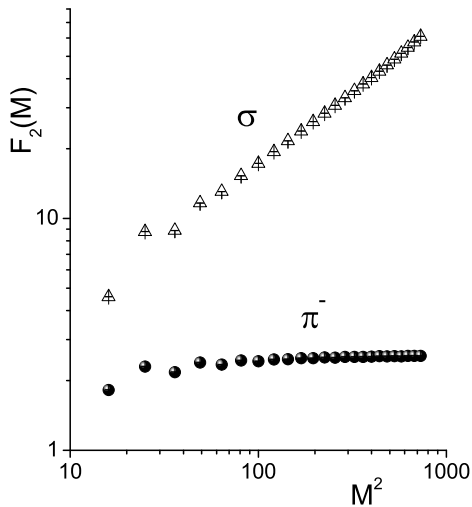}
\caption{\label{fig:fig1} The second factorial moment $F_2$ in transverse momentum space
of the negative pion sector produced through the decay of the critical sigmas in the 
$10^5$ CMC events. For comparison we
show in the same plot the second moment for the sigmas before their decay.}
\end{figure}

The reason for the
suppression of fluctuations in the pionic sector is the
kinematical distortion of the self-similar pattern formed in the 
sigma sector due to the sigma-decay. The strength of this distortion
increases with momentum transfer: $Q=\sqrt{m^2_{\sigma}-4m_{\pi}^2}$ and
becomes negligible near the two-pion threshold ($m_{\sigma} \approx 2 m_{\pi}$).
Here $m_{\sigma}^2=(p_{\pi^+}+p_{\pi^-})^2$ where $p_{\pi^{\pm}}$ are
the four momenta of the produced charged pions.

In this treatment we consider only the visible mode of the sigma-decay
to charged pions and the reconstruction of the sigmas, based on
this sector, is simply a search for $(\pi^+,\pi^-)$ pairs with invariant mass
close to $2 m_{\pi}$. The aim of this procedure is to establish a self-consistent
mechanism for producing critical sigmas either directly from the event generator
(CMC) or through reconstruction from the final states $(\pi^+,\pi^-)$ in the
same data set of simulated events. The significance of this self-consistent
reconstruction approach in a search for critical $\sigma$-fluctuations in real
events becomes self-evident.
In practice this can be done by looking, in each recorded event, for $(\pi^+,\pi^-)$ 
pairs fullfiling the criterion A:
\begin{equation}
A=\{(\pi^+,\pi^-) \vert 4 m_{\pi}^2 \leq (p_{\pi^+} + p_{\pi^-})^2 \leq 
(2 m_{\pi} + \epsilon)^2 \}
\label{eq:eq1}
\end{equation}
with $\frac{\epsilon}{2 m_{\pi}} \ll 1$.
The momentum of the sigma-particle is then obtained as:
$\vec{p}_{\sigma}=\vec{p}_{\pi^+} + \vec{p}_{\pi^-}$. 
In general the pairing will not be one-to-one, allowing 
for different possibilities to pair the charged pions in order to get 
sigmas. It turns out that the {\bf full} pairing forming {\bf all} possible pairs 
$(\pi^+,\pi^-)$ fullfiling (\ref{eq:eq1}) for a given $\pi^+$, produces more accurately
the geometric properties of the critical sigma sector in the limit $\epsilon \to 0$.
Due to finite statistics this limit can only be taken using data sets consisting of events
with reconstructed sigmas, obtained for a sequence 
of $\epsilon$-values approaching to zero, and then extrapolate smoothly to $\epsilon=0$. 
Thus we have performed factorial moment analysis in transverse momentum space
using values of $\epsilon$ in the region $8-15~MeV$. We have calculated the second 
moment $F_2$ for different sets of events with reconstructed sigmas, each obtained using a fixed 
$\epsilon$. It exists a wide window in the resolution scale $M$
for which $F_2 \sim (M^2)^{s_2(\epsilon)}$.  
Using linear extrapolation the full pairing leads to the value 
$s_2(0)=0.70 \pm 0.03$ very close to the theoretically predicted $s_2(0)=\frac{2}{3}$. 
In an alternative rule for pairing, choosing for each $\pi^+$ {\bf randomly} the
corresponding $\pi^-$ partner, from all possible ones fullfiling (\ref{eq:eq1}), and repeating
the same moment analysis we find $s_2(0) \approx 0.44 \pm 0.01$ deviating by $35 \%$ from
the theoretically expected value.
Thus the full pairing reproduces succesfully the geometrical properties of
the critical system due to the fact that in this way one reconstructs
{\bf all} critical sigmas even if it is impossible to identify them. 

The above algorithm besides the reconstruction of the real critical sigmas 
also creates a number of fake sigmas. Some of these are generated by combining pions 
belonging to different critical sigmas and contribute significantly to the strong 
intermittency effect occuring in the reconstructed 
isoscalar sector. Therefore the intermittency effect in the factorial moment analysis is a
necessary but not sufficient condition to identify critical, strongly correlated
isoscalars. There is a need to substract the effect of the fake sigmas in order to reveal the
correlated critical isoscalar sector. To achieve this goal one has to calculate the 
pair-correlation density of the sigmas defined by:
\begin{equation}
C(1,2)=<\rho_2(1,2)>-<\rho_2^{(m)}(1,2)>
\label{eq:eq2}
\end{equation}
where $<\rho_2>$ is the two-particle density averaged over the events.
The arguments $1,2$ refer to the momentum
coordinates of two sigma-particles. The second term is the two-particle density
estimated using mixed events \cite{WolfDr}. In fact for a given set of events one can 
construct a set of mixed events by shuffling the momentum coordinates of the pions 
in the initial data set.
As a consequence the (possible) correlations in the original data, including the formation
of sigmas, are 
destroyed in the mixed set. In general the two-particle density is given 
by the number of particle pairs per event and per phase space cell of volume $d \Omega$.
Therefore eq.(\ref{eq:eq2}) can be written as:
\begin{equation}
C(1,2)=\frac{1}{N_{ev}} \sum_{i=1}^{N_{ev}} \frac{d n_{12,i}}{d \Omega} - W
\frac{1}{N_{ev}^{(m)}} \sum_{i=1}^{N_{ev}^{(m)}} \frac{d n_{12,i}^{(m)}}{d \Omega} 
\label{eq:eq3}
\end{equation}
where $i$ labels the events, $N_{ev}$ is the total number of CMC events fullfiling
the condition (\ref{eq:eq1}) while $N_{ev}^{(m)}$ is the total number of mixed events,
obtained from the shuffling of the tracks in the CMC events, fullfiling (\ref{eq:eq1}). 
The coefficient $W$ is the ratio of numbers of events: $W=\frac{N_{ev}^{(m)}}{N_{ev}}$. 
The correlator $G(M)$ is defined as the integral of 
$C(1,2)$ over a finite phase space cell with volume $\delta \Omega=\frac{\Omega}{M^D}$, 
where $\Omega$ is the total available phase space in $D$-dimensions (here $D=2$). 
Performing the integration and normalizing properly we 
can express the correlator in terms of the second order factorial moments \cite{WolfDr}. 
Taking into account 
the fact that the resulting distributions of the sigmas depend on the parameter $\epsilon$ 
through (\ref{eq:eq1}) we finally get:
\begin{equation}
G(M,\epsilon)=F_2(M,\epsilon)-W(\epsilon)\left(\frac{<n_{\sigma}^{(m)}>}
{<n_{\sigma}>}\right)^2
F_2^{(m)}(M,\epsilon)
\label{eq:eq4}
\end{equation}
where $<n_{\sigma}>$ is the mean multiplicity of sigmas within a phase space cell for
the CMC events while $<n_{\sigma}^{(m)}>$ is the corresponding quantity for the
mixed events. These two multiplicities are in fact almost equal and therefore
their ratio appearing in eq.(\ref{eq:eq4}) is very close to one for all $\epsilon$. 
The weight factor $W$ has a transparent physical interpretation:
it describes the relative weight of the fake sigmas in the reconstructed set. If 
$W(\epsilon) \approx 1$ the fake sigmas dominate while for $W(\epsilon) \to 0$ they are 
suppressed relative to the real sigmas. It turns out 
that the dependence of $W$ on $\epsilon$ is crucial for the determination of the critical
isoscalar sector. Accordingly, if $W \approx 0$ the correlator follows the properties
of the second moment $F_2$ and the effect of the mixed events is marginal, while for 
$W \approx 1$ the correlator becomes negligible and no critical effect can be identified.
At threshold ($\epsilon=0$) we expect to fully recover the critical isoscalar sector 
therefore $W(0)$ must vanish if real critical sigmas exist in the data set of the reconstructed
sigmas. This should be the case when analyzing the CMC events. In fact, it turns out that in
order to avoid fake sigmas at threshold, in the reconstruction procedure, events with
sufficient multiplicity of pion pairs $(n_{\sigma})$ must be considered in the 
analysis $(n_{\sigma} > 2)$. In addition the correlator 
$G(M,\epsilon)$ should behave like $~ \sim M^{2 \phi_2(\epsilon)}$ if self-similar patterns
of correlated sigmas are present in the corresponding phase space. The exponent 
$\phi_2(\epsilon)$ is a new intermittency exponent having the property $\phi_2(0)=s_2(0)$ as
$G(M,0) \approx F_2(M,0)$. Finally, the strength of the correlations in the sigma sector is 
given by the magnitude of $G(M,\epsilon)$. Therefore in terms of the correlator we have three
very strict requirements in order to reveal the critical isoscalars 
in a data set of charged pions:
\begin{itemize}
\item{Dominance of the real sigmas with respect to the fake sigmas when approaching the 
two-pion threshold ($\lim_{\epsilon \to 0} W(\epsilon) =0$)}
\item{Large positive values of the correlator at threshold ($G(M,0) \gg 1$) at high
resolution scales.}    
\item{Correct power-law dependence of the correlator on the resolution scale $M$, at threshold:
$G(M,0) \sim M^{1.34}$}
\end{itemize}
In a conventional hadronic system with no critical isoscalar sector at least one of the
above properties will not be met.

To test the above ideas we applied the sigma-reconstruction algorithm to $10^5$ CMC events
describing a small system as well as to $33000$ HIJING events \cite{Hijing} for the
$C+C$ system at the SPS. The mean charged pion multiplicity per event in these two systems
is practically the same. The CMC data set is expected to
possess a recoverable critical isoscalar sector. On the contrary the HIJING simulator describes,
in the low $p_{\perp}$ region, conventional hadronic dynamics and 
therefore no critical isoscalars are expected to
occur in the corresponding data set \cite{Hijing}. 

We have first calculated the correlator for these two systems choosing $n_{\sigma} \geq 5$ 
for the multiplicity of pion pairs. The results are given in Fig.~2 for $\epsilon$-values 
in the region of $8-15~MeV$, suggested by finite statistics.  

\begin{figure*}
\includegraphics{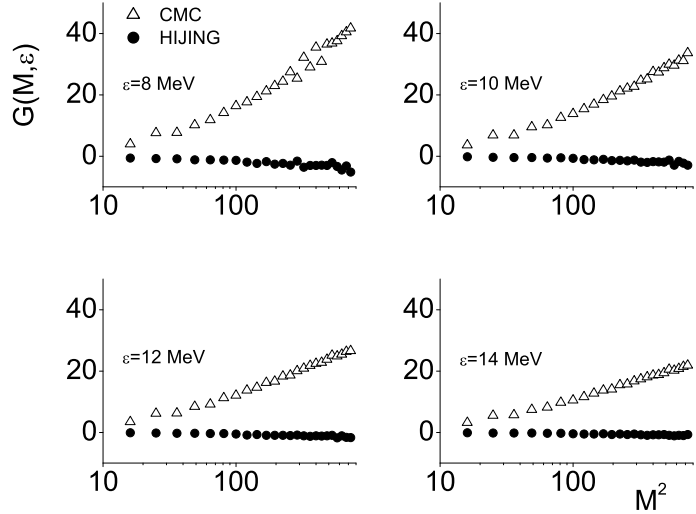}
\caption{\label{fig:fig2} The correlator $G(M,\epsilon)$ for the reconstructed 
sigmas in $10^5$ CMC events and
in 33000 HIJING events. We present the results for 4 different values of 
$\epsilon$.}
\end{figure*}

As we can see the CMC-correlator behaves very differently from
the HIJING-correlator for all values of $\epsilon$. In Fig.~3 we also show 
the weight factor $W(\epsilon)$ for both systems. It is obvious that in the CMC case $W$ 
approaches zero for $\epsilon \to 0$ while in HIJING $W(\epsilon)$ is of order of
one independently of the $\epsilon$-value. The interpretation of this very important result is 
that in the case of HIJING, i.e. in a conventional hadronic system, the reconstructed sigmas 
are fake for every value of $\epsilon$. 

\begin{figure}
\includegraphics{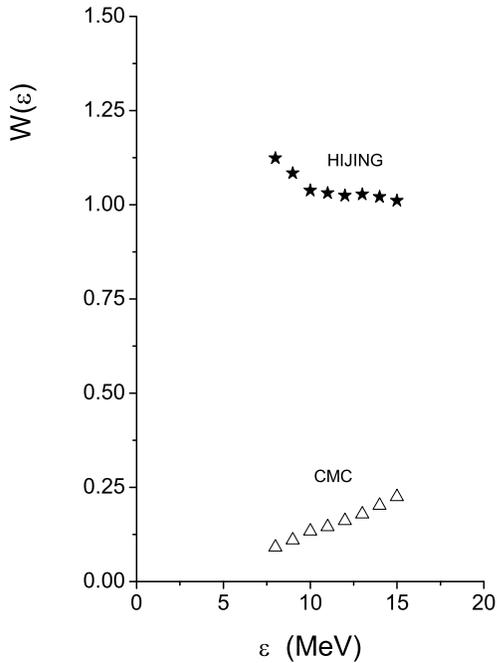}
\caption{\label{fig:fig3} The weight factor $W(\epsilon)$ for the 
CMC and the HIJING events.}
\end{figure}

Finally in
Fig.~4 we plot $\phi_2(\epsilon)$ for the CMC case. The linear extrapolation leads to 
$\phi_2(0)=0.69 \pm 0.04$ in very good agreement with the theoretically expected value of
$2/3$. According to the analysis above we have recovered in a self-consistent manner the
critical isoscalar sector in the CMC data set as a result of $\sigma$-reconstruction. 
On the contrary the HIJING events do not contain real critical sigmas, as expected.

\begin{figure}
\includegraphics{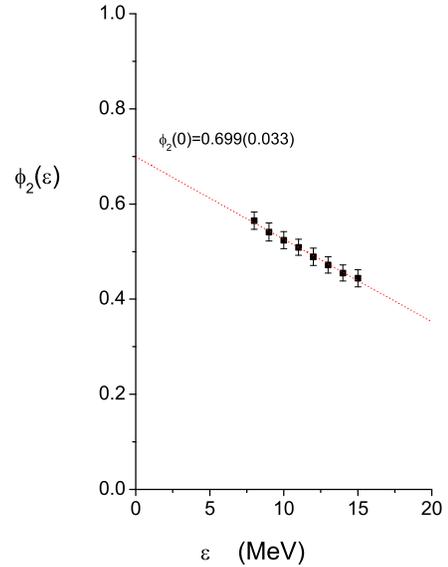}
\caption{\label{fig:fig4} The slope $\phi_2(\epsilon)$ of the correlator 
for the CMC system.}
\end{figure}

We also examined the dependence of our results on the charged pion multiplicity. We have 
produced CMC events for a larger system with size in rapidity $\Delta=11$ and $R_{\perp}=25~fm$.
In this case the total charged pion multiplicity is $\approx 120$ per event coming from the 
decay of sigmas with mean multiplicity $\approx 96$. 
We found that $W(0)$ is very close to zero (linear extrapolation: $W(0)=-0.03$) 
and $\phi_2(0)=0.664 \pm 0.02$ in very good agreement with the estimations 
in the smaller critical system (Fig.~4). Thus the results concerning the 
description of the critical 
isoscalar sector are not sensitive to the size of the system, as expected. 
What is changing is the correlation length which for a second order 
phase transition is proportional to the size of the
system. In fact for the two systems considered here the critical 
correlation length is $\approx 6~fm$ 
(small system) and $\approx 10~fm$ (large system) respectively \cite{ACDPRE}.

In order to support further our claim that in the CMC description the sector of $\pi^+ \pi^-$
pairs near two-pion threshold, has a dinstict dynamical behaviour, associated with 
$\sigma$-field fluctuations at the critical point, we have examined the behaviour of 
$\pi^+ \pi^+$ pairs in the same region $(\epsilon \stackrel{<}{\sim} 4~MeV)$. As expected
no effect is found, the pairing $\pi^+ \pi^+$ has no dynamical origin $(W \approx 1)$ and the
correlator $G(M,\epsilon)$ of these pairs is strongly suppressed. We illustrate this
in Fig.~5 where we compare, for $\epsilon = 4~MeV$, the correlators of sigmas ($\pi^+ \pi^-$ 
pairs) and $\pi^+ \pi^+$ pairs for the small system $(<n_{ch}> \approx 40)$. It must be noted
that for this value of $\epsilon$ the Coulomb correlations are negligible 
($Q^2=-(p_{\pi^+}-p_{\pi^-})^2 \approx 50~MeV$ \cite{Seyboth}).
Similar results are found if the system is larger $(<n_{ch}> \approx 120)$.

\begin{figure}
\includegraphics{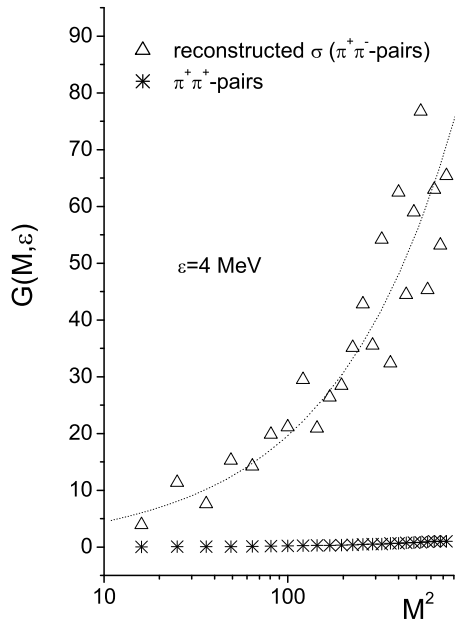}
\caption{\label{fig:fig5} The correlator $G(M,\epsilon)$ for $(\pi^+,\pi^+)$ 
pairs (stars) as well as for sigmas (open triangles) using $\epsilon=4~MeV$.}
\end{figure}

In conclusion, we have shown that a set of well prescribed observables (factorial moments,
correlators, statistical weight of fake sigmas, intermittency exponents) associated with
the existence of a critical point in quark matter, can be established in nuclear collisions.
These observables belong to the reconstructed $\sigma$-sector describing massive scalars 
($\pi^+ \pi^-$ pairs) near the two-pion threshold and their behaviour reveals strong critical
effects suggested by $\sigma$-field fluctuations near the critical point. We claim that a 
search for such a critical behaviour in heavy ion experiments is feasible within the 
framework of
a reconstruction procedure of massive scalars, discussed in this work. Since the critical
effects in this sector are strong and their pattern remains robust for
systems of different size, our proposal is to study, using the above observables, different 
processes at the SPS and RHIC with the aim 
to scan a substantial area of the phase diagram, in a systematic search for the QCD critical
point in collisions of nuclei.

\begin{acknowledgments}
We thank the NA49 Collaboration and especially K. Perl and R. Korus for their help in the
treatment of the HIJING data for the $C+C$ system. One of us (N.G.~A.) wishes to thank Larry
McLerran, Dimtri Kharzeev and Jacek Wosiek for their valuable comments. This work was partly
supported by the Research Committee of the University of Athens. 
\end{acknowledgments}

{}

\end{document}